\documentstyle[12pt]{article} 

\newcommand{\be}{\begin{equation}}
\newcommand{\ee}{\end{equation}}
\newcommand{\bea}{\begin{eqnarray}}
\newcommand{\eea}{\end{eqnarray}}
\newcommand{\bean}{\begin{eqnarray*}}
\newcommand{\eean}{\end{eqnarray*}}

\def\thefiglist#1{\section*{Figure Captions\markboth
{FIGURE CAPTIONS} {FIGURE CAPTIONS}}\list
{Figure \arabic{enumi}}
{\settowidth\labelwidth{Figure #1.}\leftmargin\labelwidth
\advance\leftmargin\labelsep
\usecounter{enumi}.}
\def\newblock{\hskip .11em plus .33em minus -.07em}
\sloppy}

\begin{document}
\title{Comment on the review paper of Amsler}
\author{D.V.Bugg$^{a}$ and A.V.Sarantsev$^{b}$\\
{\small {\it a) Queen Mary and Westfield College, London E14NS, UK}}\\
{\small {\it b) St.Petersburg Nuclear Physics Institute, Gatchina,
188350 Russia}}}
\date{}
\maketitle

\begin{abstract}
We present a dissenting view of some aspects of the work of
the Crystal Barrel collaboration.
\end{abstract}

We have found through the World Wide Web a review article Claude Amsler
has submitted to Reviews of Modern Physics and entitled ``$\bar pp$
Annihilation and Meson Spectroscopy with the Crystal Barrel'', 
HEP-EX/9708025.
We have major disagreements of both history and scientific fact concerning
this review. This page summarises the points on which we disagree.
\newline

The article has not been approved by the collaboration. 
It is the private view of Amsler. 
In our view, it misrepresents the historical sequence of discoveries.
The review presents in detail a point of view developed by Zurich members of
the collaboration. They are entitled to express their
opinions. However, it seriously misrepresents much other work within the
collaboration and fails to reference many papers and experimental facts
which conflict with the Zurich line. We appreciate that disagreements
can occur, but we believe that there is a responsiblity to expose these
disagreements in a review article. Otherwise, the reader could be misled
into believing the disagreements have been resolved in favour of the view
being expressed; that is not the case here.
\newline

One of the important results cited by Amsler is the
discovery of three $0^+$ mesons; (a) $f_0(1500)$, now regarded as the prime
candidate for the $0^+$ glueball, (b) $f_0(1300-1370)$ and (c) $a_0(1450)$.
We wish to set straight the actual history of this discovery.
There were two crucial steps.
The first was the production during 1990-1992 of data on channels $3\pi ^0$,
$\eta \pi ^0 \pi ^0$ and $\eta \eta \pi ^0$ by Mainz
and Karlsruhe groups.
The second step was a Coupled Channel analysis by QMWC and St. Petersburg
groups, using the N/D method and showing that all  these three channels 
could be
fitted consistently with the three $0^+$ resonances mentioned above.
The latter work was done in August 1992, during a
visit to Queen Mary and Westfield College by Anisovich and Sarantsev from
St. Petersburg.
Their visit was specifically in order to carry out this analysis of the
Crystal Barrel data and was cleared in advance with the Steering
Committee of the collaboration.
For the purposes of this work, they are seen officially as visitors to
Queen Mary and Westfield College.
The work was reported to the collaboration at its meeting in September 1992, as
may be verified from the minutes of the collaboration.
After much discussion and checking by both the Mainz group and by Michael 
Kobel (CERN), it was cleared for publication at the collaboration meeting 
in July 1993. The analysis was
presented at the Marseille conference at the end of that month.
It was also presented shortly afterwards at the NAN conference in Moscow.
Reference to the first two presentations are not given by Amsler.
They are:
\newline
(a) Hadron Spectroscopy, D.V. Bugg, Int. Europhysics Conf. on High Energy
Physics, Marseille, France, ed. J. Carr and M. Perrottet, (Editions Frontieres,
Gif-sur-Yvette, France, 1994), p717.
\newline
(b) V.V. Anisovich, Proc. Second Biennial Workshop on Nucleon-Antinucleon
Physics (NAN93), Phys. of Atomic Nuclei 57 (1994) 1595.
\newline
Subsequently, this Coupled-Channel analysis was published in four steps:
\newline
(c) V.V. Anisovich et al., Phys. Lett. B323 (1994) 233,
\newline
(d) C. Amsler et al., Phys. Lett. B333 (1994) 277
\newline
(e) V.V. Anisovich et al., Phys. Rev. D50 (1994) 1972
\newline
(f) D.V. Bugg et al., Phys. Rev. D50 (1994) 4412.
\newline
The last of these references is not given by Amsler. 
\newline

  What has happened since 1995 is that two competing analyses have developed
and the available statistics have increased roughly tenfold.
The Zurich group have developed a Coupled Channel
Analysis using the K-matrix approach and based only on Crystal Barrel data.
The first Zurich
Coupled Channel Analysis was submitted for publication on May 29, 1995 and
published roughly 1--1.5 years behind references (a)-(f) above. It appeared as
\newline
Amsler et al., Phys. Lett. B355 (1995) 425.
\newline      
                                       
We do not object to Amlser presenting the Zurich line, but it should be
made clear that there is a difference of opinion and there exists an
alternative treatment of the Crystal Barrel data from within the
collaboration itself.
An important paper giving the most recent analysis of the UK side of the
collaboration is that of  Abele et al., Nucl. Phys. A609 (1996) 562.
\newline
Some of our conclusions are significantly different to those of the Zurich 
group.
We include in our analysis Argonne and BNL 
data on $\pi ^- \pi ^+ \to K^0_SK^0_S$ and GAMS data on 
$\pi ^- \pi ^+ \to \pi ^0 \pi ^0$. 
For the $\pi \pi $ S-wave we have refitted the actual published moments of the 
CERN-Munich experiment.
These additional data are all fitted by the Abele et al. paper, where
they are shown to be entirely consistent with Crystal Barrel data, but
better in certain points of detail.
The Abele et al. paper finds a very
different mass, $1300 \pm 15$ MeV for what Amsler calls $f_0(1370)$.
It also comes out with a much narrower width. 
The same result was obtained in the framework of the K-matrix approach
(see refs. (h),(i) below).
Branching ratios are evaluated in the Abele et al paper in a different way 
to Zurich work and this leads to substantially different results,
which have a strong bearing on the physics interpretation placed on the
resonances by theorists.
\newline

Subsequent papers by Anisovich et al., listed below, bear strongly upon the
interpretation of Crystal Barrel data, The analysis has been extended 
using the T-matrix approach, then the K-matrix.
This work adds a considerable volume of data from other experiments,
namely CERN-Munich, GAMS, VES and data on $\pi ^-\pi ^+ \to K\bar K$ from
Brookhaven and the Argonne. An important feature of this work is that it has
removed the inconsistency between Crystal Barrel and GAMS by 
demonstrating that
the GAMS $f_0(1590)$ in explained by interference between 
a broad background amplitude and $f_0(1500)$. 
References are as follows:
\newline
(g) V.V. Anisovich, A.A. Kondashov, Yu.D. Prokoshkin,
S.A. Sadovsky and A.V. Sarantsev, Phys.Lett., B355 (1995) 363.
\newline
(h) V.V. Anisovich and A.V. Sarantsev,
HEP-PH/9603276; Phys.Lett., B382 (1996) 429.
\newline
(i) V.V. Anisovich, Yu.D. Prokoshkin and A.V. Sarantsev,
HEP-PH/9610414, Phys.Lett. 389 (1996) 388.
\newline

Apart from the $0^+$ mesons, other resonances discovered by the
collaboration have been
$a_2(1650)$, $\eta _2(1645)$ and $\eta _2(1875)$, (though there was
some earlier evidence for the latter from CELLO and Crystal Ball). New
resonances seem to us to be one of the ``achievements of the Crystal Barrel
collaboration'' and should appear in a balanced review of the experiment.
This work is not presented in the review, but is listed without comment in
the references. These resonances have both been confirmed in a recent
preprint by the Omega group at CERN, hep-ex/9707021, to be published shortly in
Physics Letters.

\end{document}